\documentclass[epj,nopacs]{svjour}
\usepackage{graphicx}
\usepackage{amssymb}
\newcommand{\nn}{\nonumber}
\def\stilde{\widetilde}
\begin{document}
\title{CP Violating Asymmetry in Stop Decay into Bottom and Chargino}
%\author{First author\inst{1} \and Second author\inst{2}% etc
\author{Helmut Eberl\thanks{\emph{e-mail:} helmut.eberl@oeaw.ac.at}
\and Sebastian M.R. Frank\thanks{\emph{e-mail:} frank@hephy.oeaw.ac.at}
\and Walter Majerotto\thanks{\emph{e-mail:} majer@hephy.oeaw.ac.at}
}                     % Do not remove
%\institute{Insert the first address here \and the second here}
\institute{Institute of High Energy Physics, Austrian Academy of Sciences, A-1050 Vienna, Austria}
\date{Received: date / Revised version: date}
\abstract{
In the MSSM with complex parameters, loop corrections to the decay of a stop into a bottom quark and a chargino can lead to a CP violating decay rate asymmetry. We calculate this asymmetry at full one-loop level and perform a detailed numerical study, analyzing the dependence on the parameters and complex phases involved. If the stop can decay into a gluino, the self-energy and the vertex correction dominate due to the strong coupling. It is shown that the vertex contribution is always suppressed. We therefore give a simple approximate formula for the asymmetry. We account for the constraints on the parameters coming from several experimental limits. Asymmetries up to 25 percent are obtained. We also comment on the feasibility of measuring this asymmetry at the LHC.
\PACS{
      {PACS-key}{discribing text of that key}   \and
      {PACS-key}{discribing text of that key}
     } % end of PACS codes
} %end of abstract
\maketitle
\section{Introduction}
If the Minimal Supersymmetric Standard Model (MSSM) is realized in nature, LHC will produce squarks and gluinos copiously. However, even if supersymmetry is discovered, it will be still a long way to determine the parameters of the underlying model. In the general MSSM, the U(1), SU(2) and SU(3) gaugino mass parameters $M_1$, $M_2$, and $M_3$, the higgsino mass parameter $\mu$, and the trilinear couplings $A_f$ (corresponding to a fermion f) may be complex, i.e. $A_f = |A_f| e^{i \varphi_{A_f}}$, $M_3 = |M_3| e^{i \varphi_{\tilde g}}$. As usual, we take $M_2$ positive and real by field redefinition \cite{Dugan:1984qf}. The experimental upper bounds on the electric dipole moment (EDM) of the electron, muon, neutron and several atoms severely constrain the phase of $\mu$. In general, the phases of $M_1$, $M_3$ and $A_{t,b}$ are much weaker constrained due to possible cancelations~\cite{Ibrahim:1997gj,Brhlik:1998zn,Bartl:1999bc,Pilaftsis:2002fe,Bartl:2003ju,Barger:2001nu,Pospelov:2005pr,Olive:2005ru,Abel:2005er,YaserAyazi:2006zw,Ellis:2008zy}. Therefore, we use real values only for $\mu$ and do not restrict the remaining phases. (For a recent discussion on EDMs see~\cite{Gajdosik:2009sd}.)

Complex MSSM parameters can lead to direct CP violation (CPV), see the summary in~\cite{Kraml:2007pr}. One example is the CP violating rate asymmetry, which is a loop induced effect. CP violating asymmetries for the production and decays of the charged Higgs $H^\pm$~\cite{Christova:2002ke,Christova:2006fb,Christova:2002sw,Christova:2003hg,Ginina:2008xa,Christova:2008bd,Christova:2008jv,Frank:2007zza,Frank:2007ca,Arhrib:2007rm} and for the decays $\tilde \chi^\pm \to W^\pm \tilde \chi^0$~\cite{Eberl:2005ay} were already studied in detail. Studies of measuring direct CP violation in stop cascade decays at the LHC based on T-odd asymmetries built from triple products were done in~\cite{Ellis:2008hq,Deppisch:2009nj,MoortgatPick:2009jy}.

In the following, we study the CP violating decay rate asymmetry of the decays $\tilde t_i \to b \, \tilde \chi^+_k$ and $\tilde t_i^* \to \bar b \, \tilde \chi^-_k$ at full one-loop level~\cite{Frank:2009re,Frank2008}. If the channel $\tilde t_i \to \tilde g \, t$ is kinematically open, the $\tilde t_1 - \tilde t_2$ self-energy and the vertex graph with $\tilde g$ exchange are expected to dominate because of the strong coupling. But we show explicitly that the vertex contribution is suppressed. All other contributions are numerically always smaller than $\sim \! 0.5 \%$ which also means that the dependence on the phase of $M_1$ is negligible. We thus give a short analytic formula for the decay rate asymmetry $\delta^{CP}$ which approximates the total one-loop result within $5 \%$ in the range above the threshold of the $\tilde t_i \to \tilde g \, t$ decay.

In order to get a large decay rate asymmetry, not only the channel into $\tilde g$ must be open, but also large phases or phase combinations of $A_t$ and $M_3$ are necessary. In addition, the stops must be rather degenerate but with a strong mixing. The dependence on $\varphi_{A_b}$ is weak because it only enters the vertex corrections.

\section{Decay Rate Asymmetry \boldmath $\delta^{CP}$}

We define the CP violating decay rate asymmetry of the decays $\tilde t_i \to b \, \tilde \chi^+_k$ and $\tilde t_i^* \to \bar b \, \tilde \chi^-_k$ as
\begin{equation}
\delta^{CP} = \frac{\Gamma^+(\tilde t_i \to b \, \tilde \chi^+_k) - \Gamma^-(\tilde t^*_i \to \bar b \, \tilde \chi^{-}_k)}
{\Gamma^+(\tilde t_i \to b \, \tilde \chi^+_k) + \Gamma^-(\tilde t^*_i \to \bar b \, \tilde \chi^{-}_k)} \, .
\label{eq:deltaCP}
\end{equation}
The one-loop decay widths can be written as
\begin{equation}
\Gamma^\pm
\propto \sum_s |\mathcal{M}^\pm_\mathrm{tree}|^2 + 2 \mathrm{Re}
\Big( \sum_s ( \mathcal{M}^\pm_\mathrm{tree} )^\dagger
\mathcal{M}^\pm_\mathrm{loop} \Big)
\end{equation}
with the matrix elements given by
\begin{eqnarray}
\mathcal{M}^+_\mathrm{tree} & = & i \, \bar u(k_1) ( B^R_+ P_R + B^L_+
P_L ) v(-k_2) \nn \, , \\ \mathcal{M}^-_\mathrm{tree} & = & i \, \bar
u(k_2) ( B^{R}_- P_R + B^{L}_- P_L ) v(-k_1) \nn \, , \\
\mathcal{M}^+_\mathrm{loop} & = & i \, \bar u(k_1) ( \delta B^R_+
P_R + \delta B^L_+ P_L ) v(-k_2) \nn \, , \\
\mathcal{M}^-_\mathrm{loop} & = & i \, \bar u(k_2) ( \delta B^R_-
P_R + \delta B^L_- P_L ) v(-k_1)
\label{dBRL}
\end{eqnarray}
with $P_{R,L} = (1 \pm \gamma^5)/2$. The tree-level couplings $B^{R,L}_+ = B^{R,L}, B^{R,L}_- = B^{R,L*}$ are defined in Appendix~\ref{sec:lagr_int}. The form factors $\delta B^{R,L}_+$ are calculated in Section~\ref{sec:contributions}. The form factors $\delta B^{R,L}_-$ can be easily obtained by conjugating all the
couplings involved.\\
Since there is no CP violation at tree level, $\delta^{CP}$ is a UV convergent quantity which means no renormalization is necessary. Furthermore, we can write $|\mathcal{M}^\pm_\mathrm{tree}|^2$ as $|\mathcal{M}_\mathrm{tree}|^2$. Assuming that the one-loop contribution is small compared to the tree level, we use the approximation
\begin{eqnarray}
\delta^{CP} & \cong & \frac{\Gamma^+ - \Gamma^-}{2 \Gamma_\mathrm{tree}} = A^{CP}_+ - A^{CP}_- \nn \, , \\
A^{CP}_\pm & = & \frac{\mathrm{Re} \big( \sum_s ( \mathcal{M}^\pm_\mathrm{tree} )^\dagger
\mathcal{M}^\pm_\mathrm{loop} \big)}{\sum_s |\mathcal{M}_\mathrm{tree}|^2}
\label{eq:deltaCP_approx}
\end{eqnarray}
with
\begin{equation}
\sum_s |\mathcal{M}_\mathrm{tree}|^2 = \Delta ( |B^R|^2 + |B^L|^2 )
- 4 m_b m_{\tilde \chi^+_k} \mathrm{Re} ( B^{R*} B^L )
\label{eq:M-tree-generic}
\end{equation}
and
\begin{eqnarray}
\mathrm{Re} \Big( \sum_s ( \mathcal{M}^\pm_\mathrm{tree} )^\dagger
\mathcal{M}^\pm_\mathrm{loop} \Big) = \Delta \, \mathrm{Re} (
B^R_\mp \delta B^R_\pm + B^L_\mp \delta B^L_\pm ) & & \nn \\ 
- 2 m_b m_{\tilde \chi^+_k} \mathrm{Re} ( B^R_\mp \delta B^L_\pm + B^L_\mp \delta B^R_\pm ) & &
\end{eqnarray}
using $\Delta = (m_{\tilde t_i}^2 - m_b^2 - m_{\tilde \chi^+_k}^2)$. By defining combined coupling matrices
\begin{equation}
C^{i j}_\pm = B^i_\mp \delta B^j_\pm
\label{eq:C-ij-pm}
\end{equation}
with $i,j = R,L$ we obtain
\begin{eqnarray}
\mathrm{Re} \Big( \sum_s ( \mathcal{M}^\pm_\mathrm{tree} )^\dagger
\mathcal{M}^\pm_\mathrm{loop} \Big) = \Delta \left( \mathrm{Re} (C^{RR}_\pm)
+ \mathrm{Re} (C^{LL}_\pm) \right) & & \nn \\
- 2 m_b m_{\tilde \chi^+_k} \left( \mathrm{Re} (C^{RL}_\pm) +
\mathrm{Re} (C^{LR}_\pm) \right) . & &
\label{eq:M-loop-generic-combined}
\end{eqnarray}
These coupling matrices can be generally expressed by $C^{i j}_\pm \propto b^i_\pm \times (g_0 g_1 g_2)^j_\pm \times \mathrm{PaVe}$ where $b^i_\pm=B^i_\mp$ is the coupling at tree level, $(g_0 g_1 g_2)^j_+$ are the couplings of the three vertices and $(g_0 g_1 g_2)^j_-$ are the conjugated couplings. PaVe stands for the Passarino--Veltman-Integrals. Omitting the indices we can write
\begin{eqnarray}
& & \mathrm{Re} (C^{i j}_\pm) \propto \mathrm{Re} ((b \, g_0 g_1 g_2)^\pm \times \mathrm{PaVe}) = \nn \\
& & \mathrm{Re} (b \, g_0 g_1 g_2) \mathrm{Re} (\mathrm{PaVe}) \mp \mathrm{Im} (b \, g_0 g_1 g_2) \mathrm{Im} (\mathrm{PaVe})
\end{eqnarray}
which leads us to the decomposition into CP invariant and CP violating parts
\begin{equation}
\mathrm{Re} (C^{i j}_\pm) = C^{i j}_\mathrm{inv} \pm \frac{1}{2}
C^{i j}_{CP} \label{eq:C_inv&C_CP}
\end{equation}
with the definitions
\begin{eqnarray}
C^{i j}_\mathrm{inv} & \propto & \mathrm{Re} (b \, g_0 g_1 g_2) \mathrm{Re} (\mathrm{PaVe}) \nn \, , \\
C^{i j}_{CP} & \propto & - 2 \mathrm{Im} (b \, g_0 g_1 g_2) \mathrm{Im} (\mathrm{PaVe}) \, .
\label{eq:definitions-C_inv&C_CP}
\end{eqnarray}
In order to obtain a non-zero $\delta^{CP}$, not only the couplings but also the PaVe's must be complex. For that at least a second decay channel must be kinematically open, i.e. a particular one-loop diagram only contributes to the asymmetry if the corresponding two-body decay is kinematically open. The asymmetry $\delta^{CP} = A^{CP}_+ - A^{CP}_-$ then becomes
\begin{eqnarray}
\delta^{CP} & = &
\Big( \Delta (C^{RR}_{CP} + C^{LL}_{CP}) -
2 m_b m_{\tilde \chi^+_k} (C^{RL}_{CP} + C^{LR}_{CP}) \Big) \nn \\
& & / \sum_s |\mathcal{M}_\mathrm{tree}|^2 \, .
\label{eq:deltaCP-C_P}
\end{eqnarray}
Neglecting the bottom mass in Eqs.~(\ref{eq:M-tree-generic},\ref{eq:M-loop-generic-combined}) the general formula of the decay rate asymmetry for a specific one-loop contribution simplifies to
\begin{equation}
\delta^{CP} \cong \frac{1}{|B^R|^2 + |B^L|^2} (C^{RR}_{CP} + C^{LL}_{CP}) \, .
\label{eq:deltaCP-C_P-simple}
\end{equation}

\noindent
Furthermore, we point out that in the asymmetry $\delta^{CP}$ possible rescattering effects (which are CP conserving) cancel each other and therefore drop out.

\section{CP Violating Contributions}
\label{sec:contributions}

\begin{figure*}[htbp]\sidecaption
\setlength{\unitlength}{1ex}
\begin{picture}(45,36)
%\put(0,0){\line(0,1){36}}
%\put(0,0){\line(1,0){45}}
%\put(0,36){\line(1,0){45}}
%\put(45,0){\line(0,1){36}}
\put(-1,7.7){\includegraphics[width=45ex]{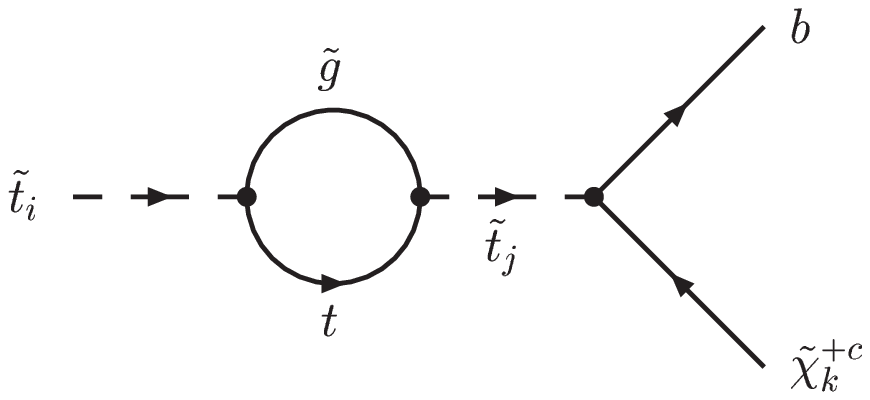}}
\end{picture}
%\hspace{2ex}
\begin{picture}(45,36)
%\put(0,0){\line(0,1){36}}
%\put(0,0){\line(1,0){45}}
%\put(0,36){\line(1,0){45}}
%\put(45,0){\line(0,1){36}}
\put(0,0){\includegraphics[width=45ex]{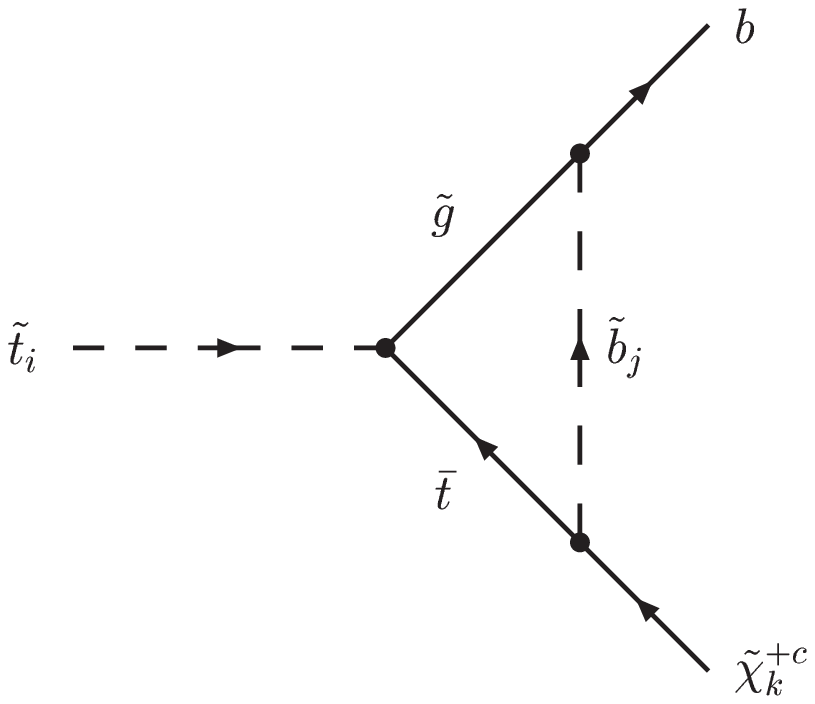}}
\end{picture}
\caption{Feynman graphs with $\tilde g$ exchange contributing to $\delta^{CP}$ of $\tilde t_i \to b \, \tilde \chi^+_k$ ($i,k = 1,2$; $j \neq i$ for the loop graph and $j = 1,2$ for the vertex correction graph)}
\label{fig:CP-violation}
\end{figure*}

In general, 47 one-loop diagrams can contribute. If the channel $\tilde t_i \to t \, \tilde g$ is kinematically open, the self-energy and the vertex graph (see Figure~\ref{fig:CP-violation}) with gluino exchange dominate due to the strong coupling.
The form factors $\delta B^{R,L}_\pm$ are defined in Eq.~(\ref{dBRL}). In the following, we only give the results for the form factor $\delta B^R_+$ since $\delta B^L_+ = \delta B^R_+ \, \mathrm{with} \, R \leftrightarrow L$. The form factor for the self-energy process is
\begin{eqnarray}
\delta B^R_+ & = & \frac{2\, C_F\, m_{\tilde g} m_t}{(4 \pi)^2
(m_{\tilde t_i}^2 - m_{\tilde t_j}^2)} B^R_{k j} \nn \\
& & \cdot \Big( G^{R *}_i G^{L}_j + G^{L *}_i G^{R}_j \Big) B_0 (m_{\tilde t_i}^2,m_{\tilde g}^2,m_t^2)
\label{eq:delta_B_R-selfenergy}
\end{eqnarray}
with $j \neq i$ and $C_F = 4/3$. The form factor for the vertex correction reads (defining $g_{i j k}^{\alpha \beta \gamma} = G^{\alpha *}_i G^{\beta *}_j A^{\gamma *}_{k j}$ with $\alpha, \beta, \gamma = R,L$)
\begin{eqnarray}
& & \delta B^R_+ = - \frac{C_F}{(4 \pi)^2} \sum_{j=1}^{2} \bigg[
g_{i j k}^{R L L} \Big( B_0(m_{\tilde t_i}^2,m_{\tilde g}^2,m_t^2) + m_{\tilde b_j}^2 C_0 \Big) \nn \\
& & + ( g_{i j k}^{L R R} m_b m_{\tilde \chi^+_k} + g_{i j k}^{L L L} m_{\tilde g} m_t \nn \\
& & + g_{i j k}^{L R L} m_b m_t + g_{i j k}^{L L R} m_{\tilde \chi^+_k} m_{\tilde g} ) C_0 \nn \\
& & + ( g_{i j k}^{R L L} m_b + g_{i j k}^{R R L} m_{\tilde g} + g_{i j k}^{L R R} m_{\tilde \chi^+_k} + g_{i j k}^{L R L} m_t ) m_b C_1 \nn \\
& & + ( g_{i j k}^{L R R} m_b + g_{i j k}^{L L R} m_{\tilde g} + g_{i j k}^{R L L} m_{\tilde \chi^+_k} + g_{i j k}^{R L R} m_t ) m_{\tilde \chi^+_k} C_2
\bigg] \, . \nn \\
\label{eq:delta_B_R-vertex}
\end{eqnarray}
The coupling matrices $A^{R,L}_{kj}$,  $B^{R,L}_{kj}$, and $G^{R,L}_i$ are given in Appendix~\ref{sec:lagr_int}. We use the one-loop integrals $B_0$, $C_0$, $C_1$, and $C_2$ according to the definition of \cite{pave} in the convention of \cite{denner}. The argument set  for the $C$-functions is $(m_b^2,m_{\tilde t_i}^2,m_{\tilde \chi^+_k}^2,m_{\tilde b_j}^2,m_{\tilde g}^2,m_t^2)$.

In order to obtain $\delta^{CP}$ we insert $\delta B^{R,L}_+$ (Eq.~(\ref{eq:delta_B_R-selfenergy}) or Eq.~(\ref{eq:delta_B_R-vertex})) into $C^{i j}_\pm$ (Eq.~(\ref{eq:C-ij-pm})). Then we calculate $C^{i j}_{CP}$ from Eq.~(\ref{eq:definitions-C_inv&C_CP}) and finally we get $\delta^{CP}$ from Eq.~(\ref{eq:deltaCP-C_P}) or Eq.~(\ref{eq:deltaCP-C_P-simple}).

We found that numerically only the self-energy graph is important. To understand the strong suppression of the gluino vertex graph in comparison to the gluino self-energy loop, one has to consider the possible combinations of the couplings in these graphs. $\delta^{CP}$ always contains a product of four squark rotation matrix elements. For the vertex graph they take the form $R^{\tilde t} R^{\tilde t*} R^{\tilde b} R^{\tilde b*}$, for the self-energy loop $R^{\tilde t} R^{\tilde t*} R^{\tilde t} R^{\tilde t*}$. Setting the indices of the external particles to be $\tilde t_1$ and $\tilde \chi^+_1$, and using the relations found in Eq.~(\ref{eq:relation_rotation_mass}), we can rewrite these terms in terms of MSSM input parameters.\\
Taking the input parameters given in Section~\ref{sec:results} we have $m_{\tilde t_1} \sim m_{\tilde t_2}$, $m_{\tilde b_1} \sim m_{\tilde b_2}$ and $\tilde \chi^+_1$ gaugino like. Since we have $m_b \, \mathrm{Im}(A_b) \ll m_t \, \mathrm{Im}(A_t)$ and assuming $\varphi_\mu \sim 0$, the remaining relevant term in $\delta^{CP}$ of the vertex graph is ($\Delta^{CP} = g_s^2 C_F m_t \mathrm{Im}(A_t) / (m_{\tilde t_2}^2 - m_{\tilde t_1}^2)$)
\begin{eqnarray}
\delta^{CP}_\mathrm{vertex} & \propto & \Delta^{CP} g^2 \frac{m_{\tilde \chi^+_1}}{m_{\tilde b_2}^2 - m_{\tilde b_1}^2} \big( (m_{\tilde b_2}^2 - m_{\tilde b_L}^2) \mathrm{Im}(C_2(1)) \nn \\
& & - (m_{\tilde b_1}^2 - m_{\tilde b_L}^2) \mathrm{Im}(C_2(2)) \big) \, ,
\end{eqnarray}
where $A_t$ is a trilinear breaking parameter, $m_{\tilde b_L}^2$ is the $(1,1)$ element of the sbottom mass matrix, and $C_2(j)$ is the Passarino--Veltman integral with $m_{\tilde b_j}$. In the limit $m_{\tilde b_1} \to m_{\tilde b_2}$ (and thus $C_2(1) \to C_2(2)$) $\delta^{CP}_\mathrm{vertex}$ vanishes. But even if the sbottom masses would not be degenerated (and thus yielding a higher numerator) the denominator always compensates this effect. The relevant term of the self-energy loop is
\begin{equation}
\delta^{CP}_\mathrm{self} \propto \Delta^{CP} g^2 \frac{m_{\tilde g}}{m_{\tilde t_2}^2 - m_{\tilde t_1}^2} \mathrm{Im}(B_0) \, .
\end{equation}
Comparing $\delta^{CP}_\mathrm{vertex}$ with $\delta^{CP}_\mathrm{self}$ we can see that the suppression of the gluino vertex correction is due to $m_b \ll m_t$, $m_{\tilde \chi^+_1} \ll m_{\tilde g}$ and  nearly degenerate sbottom masses.\\
In the case of a $\tilde \chi^+_1$ which is higgsino like, the relevant term in $\delta^{CP}_\mathrm{vertex}$ (which is now proportional to the Yukawa coupling $|h_t|^2$ instead of $g^2$) has a numerator which is again very small due to $m_{\tilde b_1} \sim m_{\tilde b_2}$ and $m_b \ll m_t$. Comparing this to the relevant term in $\delta^{CP}_\mathrm{self}$ ($g^2$ is simply replaced by $-|h_t|^2$ for $h_b \ll h_t$) one can see the same suppression mechanism at work.\\
The suppression of the vertex correction is thus a general feature, even if $\tilde \chi^+_1$ becomes a mixed state.

As a result for our scenario we can give an approximative formula for $\delta^{CP}$ for the remaining leading self-energy contribution valid for $\Gamma^\mathrm{total}_{\tilde t_j}/2 \ll |m_{\tilde t_i} - m_{\tilde t_j}|$. Inserting Eq.~(\ref{eq:delta_B_R-selfenergy}) into Eq.~(\ref{eq:C-ij-pm}) in order to calculate Eq.~(\ref{eq:definitions-C_inv&C_CP}) and finally Eq.~(\ref{eq:deltaCP-C_P-simple}) we get
\begin{eqnarray}
\delta^{CP}_\mathrm{approx} & = & - \frac{1}{4 \pi^2} \frac{C_F}{(|B^R_{k i}|^2 + |B^L_{k i}|^2)} \frac{m_{\tilde g} m_t}{(m_{\tilde t_i}^2 - m_{\tilde t_j}^2)} \nn \\
& & \cdot \mathrm{Im}(b \, g_0 g_1 g_2) \mathrm{Im}(B_0)
\label{eq:deltaCP_simple}
\end{eqnarray}
with\footnote{For the $\mathrm{Im}(B_0)$ relation see Eq.~(79) in \cite{Christova:2008jv}.}
\begin{eqnarray}
\mathrm{Im}(b \, g_0 g_1 g_2) & = & \mathrm{Im} \big[ ( B^{R *}_{k i} B^{R}_{k j} + B^{L *}_{k i} B^{L}_{k j} ) \nn \\
& & \cdot ( G^{R *}_i G^{L}_j + G^{L *}_i G^{R}_j ) \big] \nn \, , \\
\mathrm{Im}(B_0) & = & \frac{\pi \lambda^{1/2}( m_{\tilde t_i}^2, m_{\tilde g}^2, m_t^2 )}{m_{\tilde t_i}^2} \theta (m_{\tilde t_i} - m_{\tilde g} - m_t) \nn \\
\end{eqnarray}
using $\lambda (x, y, z) = x^2 + y^2 + z^2 - 2 x y - 2 x z - 2 y z$ and the step function $\theta$. $\delta^{CP}_\mathrm{approx}$ is a good approximation of $\delta^{CP}_\mathrm{all}$ (all contributions) above the threshold of the $\tilde t_i \to \tilde g \, t$ decay.

\section{Numerical Results}
\label{sec:results}

We present numerical results for the decay rate asymmetry $\delta^{CP}$ as well as the tree-level branching ratio ($BR$) of the process $\tilde t_1 \to b \, \tilde \chi^+_1$. The 47 one-loop contributions to $\delta^{CP}$ were calculated by using \mbox{\textsc{FeynArts}}~\cite{Hahn2001}. Furthermore, the gluino graphs and a few other ones were calculated independently and also cross checked numerically. The strong coupling $\alpha_s$ is taken running in the $\overline{DR}$ scheme at the scale $m_{\tilde t_1}$. In the calculation of $\delta^{CP}$ we also take the Yukawa couplings $(h_t, h_b)$ running as given in the Appendix~A of~\cite{Christova:2006fb}.\\
We calculated the EDMs up to leading two-loop order with \mbox{\textsc{CPsuperH}}~\cite{Lee:2007gn} to check that our parameter points are consistent with the constraints coming from the EDM of the electron, muon, neutron (all~\cite{Amsler:2008zzb}), and mercury~\cite{Griffith:2009zz}. We can fulfill these constraints since we take $\mu$ real, choose only the third generation breaking parameters $A_{t,b,\tau}$ to be complex, and because we can always choose the squark SUSY breaking parameters of the first and second generation appropriately. We get the right amount of the cold dark matter relic density~\cite{Komatsu:2008hk} with the $\tilde \chi^0_1$ LSP annihilating mainly into $\tau^+ \, \tau^-$ (\mbox{\textsc{micrOMEGAs}} \cite{Belanger2006,Belanger2007}) by varying $M_{\tilde L,\tilde E}$ so that $m_{\tilde \chi^0_1} \sim m_{\tilde \tau_1}$. Furthermore, the constraints coming from $B \to X_s \, \gamma$, $B_s \to \mu^+ \, \mu^-$ and $B_d \to \tau^+ \, \tau^-$ (all~\cite{Barberio:2008fa}) as well as the Higgs mass limit~\cite{Amsler:2008zzb} are fulfilled.

\begin{figure}[htbp]
\includegraphics[width=0.4\textwidth]{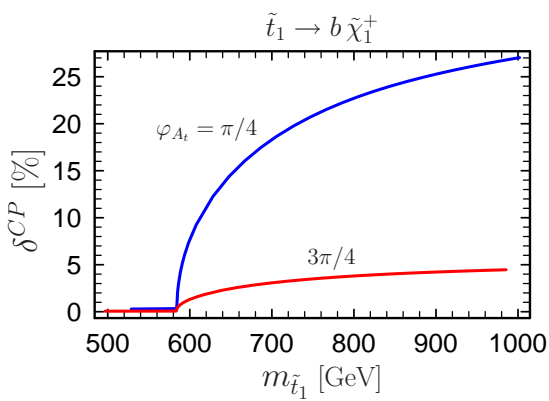}
\caption{$\delta^{CP}$ as a function of $m_{\tilde t_1}$ ($M_{\tilde Q}$ varied) for various values of $\varphi_{A_t}$}
\label{fig:delCP_mstop1_phiAt_all}
\end{figure}

\begin{figure}[htbp]
\includegraphics[width=0.4\textwidth]{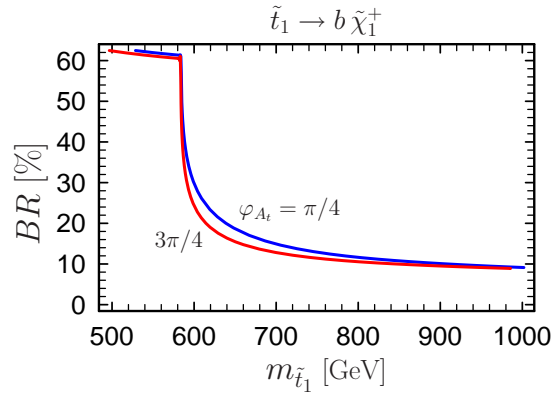}
\caption{$BR$ as a function of $m_{\tilde t_1}$ ($M_{\tilde Q}$ varied) for various values of $\varphi_{A_t}$}
\label{fig:BR_mstop1_phiAt_all}
\end{figure}

\begin{figure}[htbp]
\includegraphics[width=0.4\textwidth]{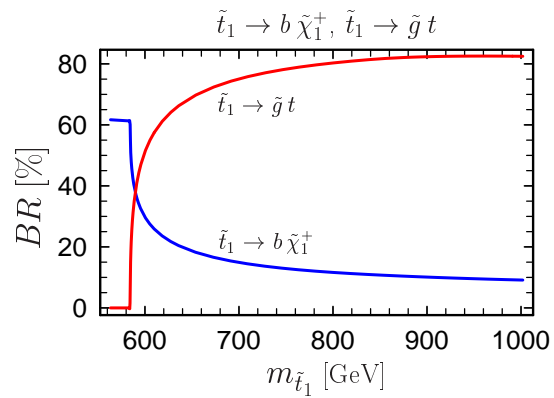}
\caption{Comparison of $BR(\tilde t_1 \to b \, \stilde \chi^+_1)$ and $BR(\tilde t_1 \to \tilde g \, t)$ as a
function of $m_{\tilde t_1}$ ($M_{\tilde Q}$ varied)}
\label{fig:BR_mstop1}
\end{figure}

For the numerical analysis, we fix for the third generation $M_{\tilde Q}=M_{\tilde U}=M_{\tilde D}$, $M_{\tilde L}=M_{\tilde E}$, $|A_t|=|A_b|=|A_{\tau}|$, $|M_1|=M_2/2$, and for the complex phases $\varphi_{A_t}=\varphi_{A_b}=\varphi_{A_{\tau}}$. We start from the following MSSM reference scenario: $M_{\tilde Q}=650$~GeV, $M_{\tilde L}=120$~GeV, $|A_t|=190$~GeV, $\varphi_{A_t}=\pi /4$, $\varphi_{\mu}=\varphi_{M_1}=\varphi_{\tilde g}=0$, $M_2=150$~GeV, $|\mu|=830$~GeV, $\tan \beta=5$, and $M_{A^0}=1000$~GeV. These parameters give $m_{\tilde \chi^+_1} = 146$~GeV, $m_{\tilde g} = 412$~GeV, $m_{\tilde t_1} = 653$~GeV, $m_{\tilde t_2} = 688$~GeV, $\Gamma^\mathrm{total}_{\tilde t_2} = 18$~GeV, $\tilde \chi^+_1$ is gaugino like, and $\tilde t_1$ and $\tilde t_2$ have a low mass splitting but large mixing.\\
In Fig.~\ref{fig:delCP_mstop1_phiAt_all} and Fig.~\ref{fig:BR_mstop1_phiAt_all} we show $\delta^{CP}$ and $BR$ as a function of $m_{\tilde t_1}$ for $\varphi_{A_t}=\frac{\pi}{4},\frac{3 \pi}{4}$. Higher values of $\varphi_{A_t}$ result in a less degenerate stop mass splitting which reduces the enhancement of $\delta^{CP}$ coming from the stop propagator (see Eq.~(\ref{eq:deltaCP_simple})). For $m_{\tilde t_1}$ the parameter $M_{\tilde Q}$ is varied from $500$ to $1000$~GeV. One can see the threshold of the $\tilde t_1 \to \tilde g \, t$ decay at $\sim \! 583$~GeV, after which the gluino contributions account for up to $\sim \! 98 \%$ of $\delta^{CP}$. However, if this decay channel opens, the $BR$ drops quickly.
This feature can be seen in detail in Fig.~\ref{fig:BR_mstop1}. If $\tilde t_1 \to \tilde g \, t$ is possible, $\delta^{CP}$ is large but $BR$ is small. The other decay channels are $\tilde t_1 \to t \, \stilde \chi^0_i , b \, \stilde \chi^+_2$.

\noindent
We also studied the dependence on the gluino phase $\varphi_{\tilde g}$ as a second source of CP violation, which is, however, in conflict with the current EDM limit of mercury by a factor of two.

\begin{figure}[htbp]
\includegraphics[width=0.4\textwidth]{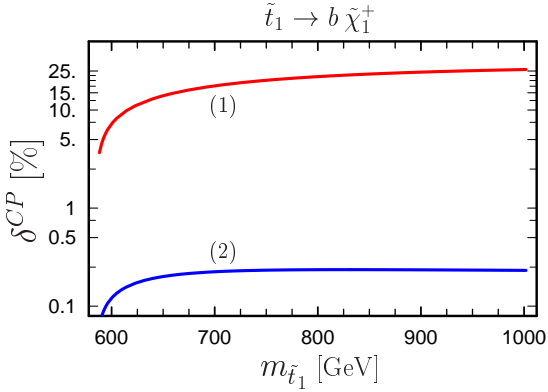}
\caption{Contribution of the gluino self-energy loop (1) and the vertex correction (2) to
$\delta^{CP}$ as a function of $m_{\tilde t_1}$ ($M_{\tilde Q}$ varied)}
\label{fig:delCP_Mstop1_gluino_compare}
\end{figure}

\noindent
Figure~\ref{fig:delCP_Mstop1_gluino_compare} shows the contributions from the self-energy and vertex graphs with gluino exchange (see Fig.~\ref{fig:CP-violation}) as a function of $m_{\tilde t_1}$.  As already anticipated in Section~\ref{sec:contributions}, the gluino self-energy loop dominates.

\begin{figure}[htbp]
\includegraphics[width=0.4\textwidth]{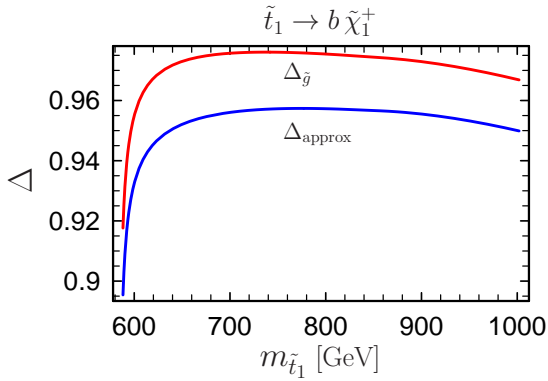}
\caption{Ratios of $\Delta_\mathrm{approx} = \delta^{CP}_\mathrm{approx} / \delta^{CP}_\mathrm{all}$,
see Eq.~(\ref{eq:deltaCP_simple}),  and $\Delta_{\tilde g} = \delta^{CP}_{\tilde g} / \delta^{CP}_\mathrm{all}$ as a function of $m_{\tilde t_1}$ ($M_{\tilde Q}$ varied)}
\label{fig:ratio_simple_gluino_loop_vs_all_and_gluino_vs_all_Mstop1}
\end{figure}

\noindent
In Fig.~\ref{fig:ratio_simple_gluino_loop_vs_all_and_gluino_vs_all_Mstop1} we show the ratios of $\delta^{CP}$ between the approximated formula for the gluino self-energy loop $\delta^{CP}_\mathrm{approx}$ (Eq.~(\ref{eq:deltaCP_simple})) and all one-loop contributions $\delta^{CP}_\mathrm{all}$, as well as between both gluino contributions $\delta^{CP}_{\tilde g}$ and all contributions. Above the threshold of $\tilde t_1 \to \tilde g \, t$ both gluino processes account for $\sim \! 97 \%$ of all processes. One can see that $\delta^{CP}_\mathrm{approx}$ is indeed a good approximation of $\delta^{CP}_\mathrm{all}$.

\noindent
We also studied the dependence of $\delta^{CP}$ on $\tan \beta$ and $|A_t|$. The main effect comes from the off-diagonal element in the stop mass matrix $a_t = A_t^* - \mu / \tan \beta$. For $|A_t| \sim 190$~GeV and $\tan \beta \sim 6$ the asymmetry has its maximum up to $25 \%$ because $a_t$ becomes minimal. In this case one has rather degenerate stop masses which enhance the gluino self-energy contribution due to the propagator $\propto 1/(m^2_{\tilde t_2} - m^2_{\tilde t_1})$. For larger values of $|A_t|$ and $\tan \beta$ the asymmetry decreases and begins to be in conflict with the EDM limit of mercury.

\begin{figure}[htbp]
\includegraphics[width=0.4\textwidth]{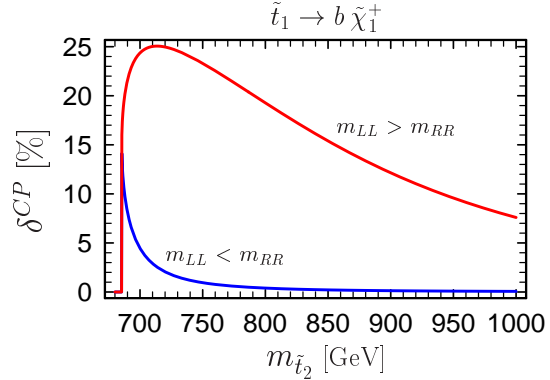}
\caption{Effect on mass splitting of $\tilde t_1$ and $\tilde t_2$ for $\delta^{CP}$ as a
function of $m_{\tilde t_2}$ ($M_{\tilde Q}, M_{\tilde U}$ varied) for $m_{\tilde t_1} = 650$~GeV}
\label{fig:delCP_mstop2_mLL_mRR}
\end{figure}

\begin{figure}[htbp]
\includegraphics[width=0.4\textwidth]{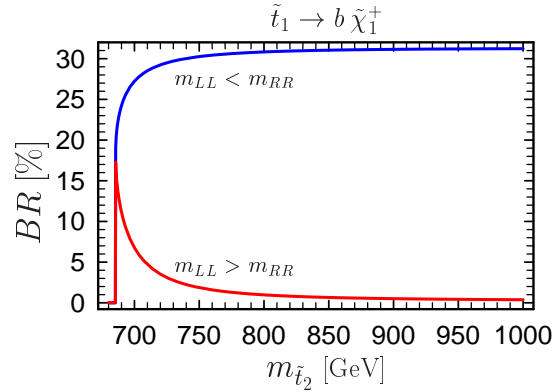}
\caption{Effect on mass splitting of $\tilde t_1$ and $\tilde t_2$ for $BR$ as a
function of $m_{\tilde t_2}$ ($M_{\tilde Q}, M_{\tilde U}$ varied) for $m_{\tilde t_1} = 650$~GeV}
\label{fig:BR_mstop2_mLL_mRR}
\end{figure}

\noindent
The effect on the mass splitting of $\tilde t_1$ and $\tilde t_2$ is shown in more detail in Fig.~\ref{fig:delCP_mstop2_mLL_mRR} and Fig.~\ref{fig:BR_mstop2_mLL_mRR}. We fix $m_{\tilde t_1} = 650$~GeV and vary $m_{\tilde t_2}$ by changing the parameters $M_{\tilde Q}, M_{\tilde U}$. There are two possibilities, $m_{LL} \lessgtr m_{RR}$.
In our scenario $\tilde \chi^+_1$ is gaugino-like which couples dominantly to $\tilde t_L$. For $m_{L L} < m_{R R}$ we have $\tilde t_1 \sim \tilde t_L$ and $\tilde t_2 \sim \tilde t_R$. Hence $BR (\tilde t_1 \to b \, \tilde \chi^+_1)$ is large, but $\delta^{CP}$ is small because the coupling of the internal $\tilde t_2$ to $b \, \tilde \chi^+_1$ is suppressed (see left graph of Figure~\ref{fig:CP-violation}). For $m_{L L} > m_{R R}$ one has the opposite behavior. When $\tilde \chi^+_1$ is higgsino-like the whole situation is reversed. We therefore see that for the cases where $\delta^{CP}$ is large the BR is always small and vice versa, unless the masses of $\tilde t_1$ and $\tilde t_2$ are rather degenerate. Note that the masses cannot be arbitrarily degenerate, since otherwise the off-diagonal elements in the stop mass matrix (and thus the complex $A_t$ parameter as the main source of CP violation) would need to vanish.\\
Furthermore, it is crucial that the enhancement of $\delta^{CP}$ due to a small stop mass difference is not a resonance enhancement, i.e. the mass difference $m_{\tilde t_2} - m_{\tilde t_1}$ must be sufficiently large compared to the FWHM $\Gamma^\mathrm{total}_{\tilde t_2}/2$ of the resonance. In our scenario this is always the case, since $\Gamma^\mathrm{total}_{\tilde t_2}/2 = 9$~GeV and $\Delta m_{\tilde t} = 35$~GeV at the reference point and $\Gamma^\mathrm{total}_{\tilde t_2}/2$  always remains much smaller for various mass differences.

\noindent
For completeness, we also examined the process $\tilde t_2 \to b \, \tilde \chi^+_1$. Because of the similar masses and large mixing of the stops, the resulting plots are alike. We obtain $\delta^{CP} = 19 \%$ and $BR = 17 \%$ at the reference point.

\noindent
We also investigated the influence on the Yukawa couplings $h_t$ and $h_b$ taken to be running. In our scenario, the difference of the asymmetry $\delta^{CP}$ taken with running and non-running Yukawa couplings is negligible.

\noindent
The theoretical uncertainty of the asymmetry $\delta^{CP}$ is estimated $\sim \! 20 \%$, due to higher order corrections~\cite{Kraml:1996kz,Guasch:2003ig}.

\noindent
For a measurement of the asymmetry $\delta^{CP}$ (Eq.~(\ref{eq:deltaCP})) at LHC, one has to consider the process
\begin{eqnarray}
p p & \to & {\tilde t}_1 \bar{\tilde t}_1 + X \to (b \tilde \chi_1^+) (\bar b \tilde \chi_1^-) + X \nn \\
& \to & (b W^+ \tilde \chi_1^0) (\bar b W^- \tilde \chi_1^0) + X \nn \\
& \to & \textrm{2 b-jets} + W^+ + W^- + \not\!\!E_{T} + X
\label{eq:signa}
\end{eqnarray}
(where $X$ only contains the beam jets) with one $W^+$ decaying hadronically and the other one leptonically to get information on the charge of the $W$, and in turn of the chargino and the stop. We assume that the masses of ${\tilde t_1}$, ${\tilde t_2}$, $\tilde \chi_1^\pm$ and $\tilde \chi_1^0$ are known so that one has enough kinematical constraints to single out the respective decays~\cite{MoortgatPick:2010wp}. In such a way it should also be possible to separate off the production and decay of ${\tilde t_1}$ from those of ${\tilde t_2}$. For $\sqrt{s}=14$~TeV, $m_{\tilde t_1}=610$~GeV the cross section of $p p \to {\tilde t}_1 \bar{\tilde t}_1$ is $\sigma=200$~fb at NLO according to \textsc{Prospino}~\cite{Beenakker1998}. Assuming ${\cal L}=300 \, \mathrm{fb}^{-1}$ and $BR (\tilde t_1 \to b \, \tilde \chi^+_1) = 0.2$ we estimate a purely statistical relative error of the asymmetry $\Delta \delta^{CP} / \delta^{CP} = \pm 0.077$. The largest background is of course pair production of top quarks~\cite{Dydak1996}. For a realistic estimate of the measurability a Monte Carlo study would be necessary, which is, however, beyond the scope of this article. For the same decay chain of ${\tilde t_1}$ such a study was done for an $e^+ e^-$ collider~\cite{Bartl:2009rt}.

\section{Conclusions}

In the MSSM with complex parameters, loop corrections to the $\tilde t_i \to b \, \stilde \chi^+_k$ decay can lead to a CP violating decay rate asymmetry $\delta^{CP}$. We studied this asymmetry at full one-loop level, analyzing the dependence on the parameters and phases. Below the threshold of the $\tilde t_i \to \tilde g \, t$ decay, $\delta^{CP}$ is $< \! 1 \%$. If this channel is open, a $\delta^{CP}$ up to $25 \%$ is possible (mainly due to the gluino contribution in the self-energy loop), if the stop particles have a small mass splitting together with large mixing and the chargino is wino like.

\begin{acknowledgement}
S.F. would like to thank Sabine Kraml for helpful correspondence. The authors acknowledge support from EU under the MRTN-CT-2006-035505 network programme. This work is also supported by the "Fonds zur F\"orderung der wissenschaftlichen Forschung" of Austria, project No.~P18959-N16.
\end{acknowledgement}

\appendix

\section{Masses and Mixing Matrices}
\label{sec:masses}

The sfermion mass matrix in the basis $(\tilde f_L,\tilde f_R)$ with $\tilde f = \tilde t, \tilde b, \tilde \tau, \dots$ is
\begin{equation}
  {\cal M}_{\tilde f}^2 = \left(
  \begin{array}{cc}
    m_{\tilde f_L}^2 & m_f a_f \\
    m_f a_f^* & m_{\tilde f_R}^2
  \end{array} \right)
\label{eq:mass-matrix-sfermion}
\end{equation}
with the following entries
\begin{eqnarray}
m_{\tilde f_L}^2 & = & M_{\lbrace \tilde Q; \tilde L \rbrace}^2 + m_f^2 \nn \\
& & + m_Z^2 \cos 2 \beta ( I_f^{3 L} - e_f \sin^2 \theta_W ) \, , \\
m_{\tilde f_R}^2 & = & M_{\lbrace \tilde U; \tilde D; \tilde E \rbrace}^2 + m_f^2 + m_Z^2 \cos 2 \beta \, e_f \sin^2 \theta_W \, , \\
m_f a_f & = & \biggl\{ \begin{array}{l}
m_u ( A_u^* - \mu \cot \beta ) \; \dots \; \textnormal{up-type sfermion} \nn \, , \\
m_d ( A_d^* - \mu \tan \beta ) \; \dots \; \textnormal{down-type sfermion} \, .
\end{array} \\
\label{eq:offdiag-mass-matrix-sfermion}
\end{eqnarray}
For the stops ($\tilde f = \tilde t$) we simply insert the corresponding values $I_t^{3 L}=1/2$, $e_t=2/3$, $M_{\tilde Q}$, $M_{\tilde U}$, $m_t$ and $A_t$. Analogously, for the sbottoms
($\tilde f = \tilde b$) we have $I_b^{3 L}=-1/2$, $e_b=-1/3$, $M_{\tilde Q}$, $M_{\tilde D}$, $m_b$ and $A_b$.\\
${\cal M}_{\tilde f}^2$ is diagonalized by the rotation matrix $R^{\tilde f}$ such that
$R^{\tilde f} {\cal M}_{\tilde f}^2 \, (R^{\tilde f})^{\dagger} = {\rm diag}(m_{\tilde f_1}^2, m_{\tilde f_2}^2)$ and ${ \tilde f_1 \choose \tilde f_2 } = R^{\tilde f} { \tilde f_L \choose \tilde f_R }$.
We have
\begin{equation}
R^{\tilde f} = \left( \begin{array}{cc}
R^{\tilde f}_{1 L} & R^{\tilde f}_{1 R} \\
R^{\tilde f}_{2 L} & R^{\tilde f}_{2 R}
\end{array} \right)
= \left( \begin{array}{cc}
\cos \theta_{\tilde f} & e^{i \varphi_{\tilde f}} \sin \theta_{\tilde f} \\
- e^{- i \varphi_{\tilde f}} \sin \theta_{\tilde f} & \cos \theta_{\tilde f}
\end{array} \right).
\label{eq:rotation-sfermion}
\end{equation}
Using the unitarity property of the sfermion rotation matrices and the diagonalization equation for the sfermion mass matrix,
 \begin{equation}
  {\cal M}_{\tilde f}^2 = \left(
  \begin{array}{cc}
    m_{\tilde f_L}^2 & m_f a_f \\
    m_f a_f^* & m_{\tilde f_R}^2
  \end{array} \right) = (R^{\tilde f})^\dagger \left(
  \begin{array}{cc}
    m_{\tilde f_1}^2 & 0                 \\
    0                & m_{\tilde f_2}^2
  \end{array} \right) R^{\tilde f} \, , 
\end{equation}
one can derive the following relations (we define $\Delta^{\tilde f} = 1 / ( m_{\tilde f_2}^2 - m_{\tilde f_1}^2 )$):
\begin{eqnarray}
R^{\tilde f}_{2 2} R^{\tilde f*}_{2 1} & = & - R^{\tilde f}_{1 2} R^{\tilde f*}_{1 1} = m_f a_f \Delta^{\tilde f} \nn \, , \\
R^{\tilde f}_{2 1} R^{\tilde f*}_{2 2} & = & - R^{\tilde f}_{1 1} R^{\tilde f*}_{1 2} = m_f a_f^* \Delta^{\tilde f} \nn \, , \\
R^{\tilde f}_{1 1} R^{\tilde f*}_{1 1} & = & R^{\tilde f}_{2 2} R^{\tilde f*}_{2 2} = ( m_{\tilde f_2}^2 - m_{\tilde f_L}^2 ) \Delta^{\tilde f} \nn \, , \\
R^{\tilde f}_{1 2} R^{\tilde f*}_{1 2} & = & R^{\tilde f}_{2 1} R^{\tilde f*}_{2 1} = - ( m_{\tilde f_1}^2 - m_{\tilde f_L}^2 ) \Delta^{\tilde f} \, .
\label{eq:relation_rotation_mass}
\end{eqnarray}

\noindent
The chargino mass matrix in the basis $(-i \lambda^+,\psi^1_{H_2})$ is
\begin{equation}
  {\cal M}_C =
  \left( \begin{array}{cc}
    M_2 & \sqrt 2 \, m_W \sin\beta \\
    \sqrt 2 \, m_W \cos\beta & \mu
  \end{array} \right).
\end{equation}
It is diagonalized by the two unitary matrices $U$ and $V$
\begin{equation}
  U^* {\cal M}_C V^{\dagger} = {\rm diag}(m_{\tilde \chi^{\pm}_1},m_{\tilde \chi^{\pm}_2}) \, , 
\end{equation}
where $m_{\tilde \chi^{\pm}_{1,2}}$ are the masses of the physical chargino states.

\section{Interaction Lagrangian}
\label{sec:lagr_int}

In this section we give the parts of the interaction Lagrangian that we need for the calculation of the leading contributions. The chargino-squark-quark interaction is described by
\begin{eqnarray}
{\cal L}_{\tilde \chi^+ q \tilde q'} & = & \bar t ( A^R_{k i} P_R + A^L_{k i} P_L ) \tilde \chi^+_k \, \tilde b_i \nn \\
& & + \bar b (B^R_{k i} P_R + B^L_{k i} P_L ) \tilde \chi^{+ c}_k \, \tilde t_i \nn \\
& & + \overline{\tilde \chi^+_k} ( A^{L *}_{k i} P_R + A^{R *}_{k i} P_L ) t \, \tilde b^*_i \nn \\
& & + \overline{\tilde \chi^{+ c}_k} ( B^{L *}_{k i} P_R + B^{R *}_{k i} P_L ) b \, \tilde t^*_i \, .
\end{eqnarray}
The couplings are
\begin{eqnarray}
A^R_{k i} & = & h^*_b R^{\tilde b *}_{i 2} U_{k 2} - g R^{\tilde b *}_{i 1} U_{k 1} \nonumber \\
& = & \frac{g}{\sqrt 2} \left( \frac{m_b}{m_W \cos \beta} U_{k 2} R^{\tilde b *}_{i 2} - \sqrt 2 U_{k 1} R^{\tilde b *}_{i 1} \right) \nonumber \, , \\
A^L_{k i} & = & h_t R^{\tilde b *}_{i 1} V^*_{k 2} = \frac{g m_t}{\sqrt 2 m_W \sin \beta} V^*_{k 2} R^{\tilde b *}_{i 1} \nonumber \, , \\
B^R_{k i} & = & h^*_t R^{\tilde t *}_{i 2} V_{k 2} - g R^{\tilde t *}_{i 1} V_{k 1} \nonumber \\
& = & \frac{g}{\sqrt 2} \left( \frac{m_t}{m_W \sin \beta} V_{k 2} R^{\tilde t *}_{i 2} - \sqrt 2 V_{k 1} R^{\tilde t *}_{i 1} \right) \nonumber \, , \\
B^L_{k i} & = & h_b R^{\tilde t *}_{i 1} U^*_{k 2} = \frac{g m_b}{\sqrt 2 m_W \cos \beta} U^*_{k 2} R^{\tilde t *}_{i 1} \, , 
\end{eqnarray}
where we used the relations $m_t = ( h_t \, \nu \sin \beta ) / \sqrt 2$, $m_b = ( h_b \, \nu \cos \beta ) / \sqrt 2$ and $m_W = g \nu /2$ to convert the coefficients.\\
The gluino-squark-quark interaction is
\begin{eqnarray}
{\cal L}_{\tilde g q \tilde q} & = & \overline{\tilde g_{\alpha}} ( G^{R \alpha}_{i;u v} P_R + G^{L \alpha}_{i;u v} P_L ) q^u \tilde q^{\, v *}_i \nn \\
& & + \overline{q^u} ( G^{L \alpha *}_{i;u v} P_R + G^{R \alpha *}_{i;u v} P_L ) \tilde g_{\alpha} \, \tilde q^{\, v}_i \, , 
\end{eqnarray}
where we used $u,v$ as the color index ($u,v=1,2,3$), $i$ as the mass index ($i=1,2$) and $\alpha$ as the gluon/gluino index ($\alpha=1,\ldots,8$). The couplings are
\begin{eqnarray}
G^{R \alpha}_{i;u v} & = & T^{\alpha}_{u v}  G^{R}_{i} \quad {\rm with} \
G^{R}_{i} = \sqrt{2} \, g_s R^{\tilde q}_{i 2} \, e^{i \frac{\varphi_{\tilde g}}{2}} \nonumber \, , \\
G^{L \alpha}_{i;u v} & = & T^{\alpha}_{u v}  G^{L}_{i}  \quad {\rm with} \
G^{L}_{i} = - \sqrt{2} \, g_s R^{\tilde q}_{i 1} \, e^{- i \frac{\varphi_{\tilde g}}{2}}
\end{eqnarray}
with $T^{\alpha}_{u v}$ as the generator of the $\mathrm{SU(3)_C}$ group, $g_s$ as the coupling constant of the strong interaction and $\varphi_{\tilde g}$ as the gluino mass phase\footnote{We agree with the gluino-squark-quark coupling given in the FeynArts Model file {\tt MSSMQCD.mod} by taking ${\tt SqrtEGl} = e^{i \frac{\varphi_{\tilde g}}{2}}$.}.

% ***************************************
% BibTex output file (.bbl) of document inserted here due to LaTeX processing issues with arXiv
% ***************************************
%
% BibTeX users please use
%\bibliographystyle{h-physrev5} % Dieser style ist noch nicht optimal (verschluckt "[hep-ph]" o.ä. bei neueren arXiv-IDs)
%\bibliography{paper}

%\begin{thebibliography}{}
%Author, \textit{Book title} (Publisher, place year) page numbers
% etc
%\end{thebibliography}

\end{document}